\documentclass[%
 aip,
 amsmath,amssymb,
 reprint,%
]{revtex4-1}

\usepackage{graphicx}% Include figure files
\usepackage{dcolumn}% Align table columns on decimal point
\usepackage{bm}% bold math
\usepackage{mathtools}
\usepackage[utf8]{inputenc}
\usepackage[T1]{fontenc}
\usepackage{mathptmx}

\DeclarePairedDelimiter\abs{\lvert}{\rvert}%

\begin{document}

\preprint{AIP/123-QED}

\title{Dynamic percolation of ferromagnetic regions in phase separated manganites using non-uniform electric fields}
% Force line breaks with \\

\author{Ambika Shakya}
 \author{Amlan Biswas}%
 \email{amlan@phys.ufl.edu}
\affiliation{ 
Department of Physics, University of Florida, Gainesville, FL 32611
}%

\date{\today}% It is always \today, today,

\begin{abstract}
Thin films of the manganite (La$_{1-y}$Pr$_y$)$_{1-x}$Ca$_x$MnO$_3$ exhibit dynamic phase coexistence with micrometer scale, fluid-like ferromagnetic metallic (FMM) regions interspersed in a charge-order insulating (COI) background. It has been previously reported that a uniform electric field realigns the fluid-like FMM regions due to a phenomenon similar to dielectrophoresis. Here we report that non-uniform electric fields have a stronger effect on the FMM regions as expected from the dielectrophoresis model. The dynamic percolation of the FMM regions is observed over a wider range of temperatures compared to the results in a uniform field. Additionally, in a non-uniform electric field, the time required for dynamic percolation along the magnetic hard axis ($t_{\mathrm{B}}$) decreased with increasing applied voltage ($V_{\mathrm{A}}$) as a power law, $V_{\mathrm{A}}^{-\delta}$ with $\delta \approx 5$ while $\delta < 2$ for a uniform electric field. Our results in a non-uniform electric field provide strong evidence in favor of the dielectrophoresis model and a unique method for manipulating micrometer-sized ferromagnetic regions using electric fields. 
\end{abstract}

\maketitle

\section{\label{sec:level1}Introduction}

The quest for electric control of magnetism has been driven by the potential for applications in spintronic devices and aided by a deeper understanding of magnetic phenomena such as multiferroism and 2-dimensional (2D) ferromagnetism~\cite{Ramesh,Cheong,Scott,Scholl,Gong,Matsukura}. $AB$O$_3$ perovskite oxides hold a significant place in spintronic device applications due to advanced film growth techniques, stability in ambient conditions, and properties such as high spin polarization~\cite{Rondinelli,Soulen}. However, electrically controlled magnetism in such oxides is usually observed in a smaller subset of non-centrosymmetric materials~\cite{Ramesh,Cheong,Scott,Scholl,Spaldin}. Here, we report a method for electric control of ferromagnetic regions which exploits the phase competition and coexistence in perovskite manganese oxides (manganites). 

Phase competition is a characteristic of manganites which, in the presence of disorder and/or strain, may lead to a mixed phase state that can be manipulated using external parameters, such as magnetic field, electric field, and strain~\cite{Jeen,Zhang,Uehara,Dhakal}. While disorder leads to phase separation with spatially static phases, long range strain leads to a dynamic (fluid) phase separated (FPS) state~\cite{Dagotto,Ahn}. In the FPS state, the competing phases behave in a fluid-like manner since the interface between the phases is not pinned due to disorder~\cite{Bishop,Zhang,Fath}. Micrometer scale dynamic phase separation in (La$_{1-y}$Pr$_y$)$_{0.67}$Ca$_{0.33}$MnO$_3$ (LPCMO) thin films was shown conclusively in a study using low temperature magnetic force microscopy in which, the evolution of the fluid-like ferromagnetic metallic (FMM) phase in an insulating matrix was observed as a function of temperature. It had been predicted that the FPS state may behave like an ``electronic soft matter" and be highly susceptible to external perturbations such as strain and electric field~\cite{Littlewood}. Following these predictions, experiments showed various forms of electric field effects on the transport properties of LPCMO~\cite{Budhani,Munakata,Dhakal}. However, these experiments were mainly based on the inhomogeneous transport properties in the phase separated state and did not exploit the dynamic behavior observed in the FPS state. A unique behavior of the FPS state, brought on by the dynamic behavior in an electric field, was first reported by Jeen {\em et al.} who observed dynamic percolation of the fluid-like FMM regions along the axis of the electric field ~\cite{Jeen}. This realignment of the FMM regions resulted in anisotropic in-plane resistance in microstructures of LPCMO grown on (110) NdGaO$_3$ (NGO) substrates. The substrate plays a central role in such experiments which rely on the dynamic behavior of the phase separated state. (110) NGO is nearly lattice matched to LPCMO which allows formation of the FPS state while dynamic phase separation was absent in LPCMO films grown on SrTiO$_3$ (STO)~\cite{Shen,Shen2}. Hence, fluid-like behavior and electric field driven realignment of the FMM regions was not observed  in the films grown on STO, as observed using magnetic force microscopy~\cite{Shen3}.

A phenomenon similar to dielectrophoresis was suggested as the reason behind the electric field induced alignment of the FMM regions ~\cite{Jeen,Shuai}. Dielectrophoresis is the mechanism by which neutral particles suspended in a fluid medium experience a net force in a non-uniform electric field~\cite{Pohl}. In the first order, the dielectrophoretic force is due to the dipole moment $\boldsymbol{p}$ induced in a neutral particle by the external electric field $\boldsymbol{E}$ and is given by the equation: 
\begin{equation}
\boldsymbol{F}_{\mathrm{DEP}} = (\boldsymbol{p}\cdot\nabla)\boldsymbol{E} = \frac{1}{2} \alpha v\nabla\abs{\boldsymbol{E}}^2
\end{equation} 
where $\alpha$ is the polarizability and $v$ is the volume of the particle~\cite{Pohl}. For a collection of suspended neutral particles, even a {\em uniform} external field generates a force between the particles due to the induced dipole moments resulting in flocculation of the particles along the electric field lines~\cite{Vankleef}. Due to the fluid-like properties of the FPS state in LPCMO, the FMM regions behave like the neutral particles suspended in a neutral medium which leads to dynamic percolation of the FMM regions even in a uniform applied electric field. The percolation time is observed to decrease at higher electric fields as is expected from the dielectrophoresis model. However, this effect was observed over a small temperature range of about 2 K ~\cite{Jeen}. A non-uniform applied field is expected to have a stronger dielectrophoretic effect due to higher order induced moments in the FMM regions and their interactions with increasing order electric field derivatives~\cite{Nili}. In this paper, we report the effect of a non-uniform applied electric field on the dynamic phase separated state in LPCMO. Microfabricated electrodes in two configurations were deposited on an LPCMO thin film to produce the non-uniform electric field. The time dependence of resistance and dynamic percolation is observed over a temperature range of about 20 K. The percolation time decreased with increasing electric field magnitudes as a power law with a higher exponent compared to the case of a uniform field. Hence, our results confirm the contribution of higher order induced moments and field derivatives to the dielectrophoresis of the FMM regions compared to previous observations in a uniform applied field. Our results provide a method for rearranging ferromagnetic regions in a material using electric fields at even lower current densities compared to magnetic skyrmions~\cite{Yu}.

\begin{figure}
\includegraphics[width=\columnwidth]{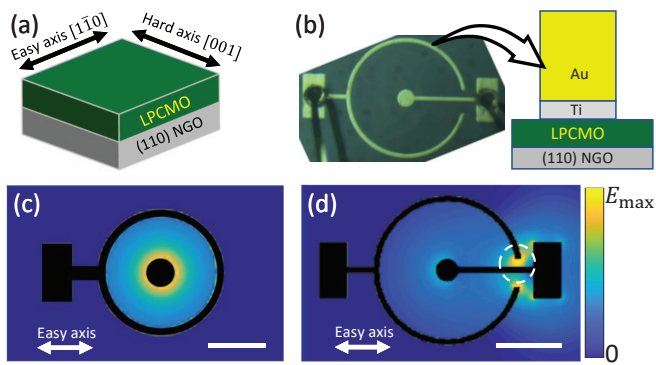}
\caption{\label{sample} (a) The orientation of the (La$_{0.4}$Pr$_{0.6}$)$_{0.67}$Ca$_{0.33}$MnO$_3$ (LPCMO) thin film grown on (110) NdGaO$_3$ (NGO) substrate shown schematically along with the directions of the magnetic easy and hard axes. (b) Optical microscope image of a microstructured gold-electrode (pattern 2) on the LPCMO thin film after lift-off photolithography and gold-wire bonding and a schematic showing a cross section of the Au \& Ti contact on the LPCMO/NGO thin film. (c) and (d) Calculated non-uniform electric field magnitudes  produced by patterns 1 and 2, respectively. The electrodes are shown in black. Scale bars in (c) and (d) correspond to 200 $\mu$m and 250 $\mu$m, respectively. The direction of the easy axis is shown for both patterns.}
\end{figure}

\section{Experimental methods}
Pulsed laser deposition technique was used to grow (La$_{0.4}$Pr$_{0.6}$)$_{0.67}$Ca$_{0.33}$MnO$_3$ (LPCMO) thin films of thickness 30 nm on (110) NdGaO$_3$ (NGO) substrates. The details of thin film growth are described in Ref. \onlinecite{Jeen2}. (110) NGO exerts anisotropic tensile strain on LPCMO of +0.49\% in the [1$\overline{1}$0] direction and +0.26\% in the [001] direction~\cite{Jeen2}. This anisotropic strain leads to in-plane magnetic anisotropy in the LPCMO film with the [1$\overline{1}$0] direction being the magnetic easy axis (see Fig.~\ref{sample}(a))~\cite{Jeen2}. The magnetic anisotropy and length scale of phase separation in LPCMO (up to about 10 $\mu$m ~\cite{Zhang}) were primary considerations in the design of the electrode patterns for applying non-uniform electric fields. The patterns were created to apply the electric field at a length scale about an order of magnitude higher than the phase separation scale i.e., about 100 $\mu$m. Such a length scale will allow us to observe the effects of dielectrophoresis without the averaging effect which occurs at millimeter scales ~\cite{Jeen}. The contribution of the magnetic anisotropy to the pattern design will be discussed in the next section. 

We deposited electrical contacts of two different shapes directly on the surface of the LPCMO thin films using lift-off photolithography process. We used a Karl Suss Delta 80 spinner to deposit positive photoresist AZ1512 of thickness 800 nm and then kept it in an oven at 105$^{\circ}$C for 25 min. UV light of wavelength 365 nm and an MA6 mask aligner was used to illuminate the photoresist through an optical mask on the sample for 40 sec in hard contact mode. Then image reversal was performed to convert the UV exposed area of positive photoresist into negative photoresist. The sample was then exposed to UV light for 60 sec in flood mode and developed for 1 minute 30 second using AZ 300 MIF developer so that the negative photoresist remained while the positive photoresist was washed away. The sample was then washed in DI water for 30 seconds. Using a KJL sputter deposition system, 20 nm Ti and 100 nm Au were deposited on the film. Finally, by cleaning with acetone and ethanol for 5 minutes in a sonicator, the metal covered photoresist layer was removed, leaving gold contact pads on the sample as shown in Fig. ~\ref{sample}(b). Note that this fabrication method does not involve any chemical etching process of the manganite thin film unlike in Refs. \onlinecite{Jeen,guneeta1}. Hence, this process avoids the formation of defects at the edges of the manganite microstructures which can act as pinning centers for the FMM regions. The two electrode patterns discussed in this paper are shown in Fig. ~\ref{sample}(c) and (d). The geometry of the patterns generate a non-uniform electric field in the plane of the thin film sample when a voltage is applied across the two (black) electrodes in Figs. 1(c) and (d). The sample was mounted on a chip carrier and gold wires bonded to the rectangular/circular pads on the sample were used for electrical contacts. A two-probe method using a constant voltage source was used to measure the resistance of the sample with these electrode configurations ~\cite{Dhakal}. The contribution of contact resistance was negligible due to the high resistances being measured which was confirmed by comparing the measured resistances with four-probe values as will be shown in the next section. The two-probe method allows us to use constant voltages across the sample for the resistance versus temperature ($R-T$) and resistance versus time ($R-t$) measurements.

\begin{figure}
\includegraphics[width=\columnwidth]{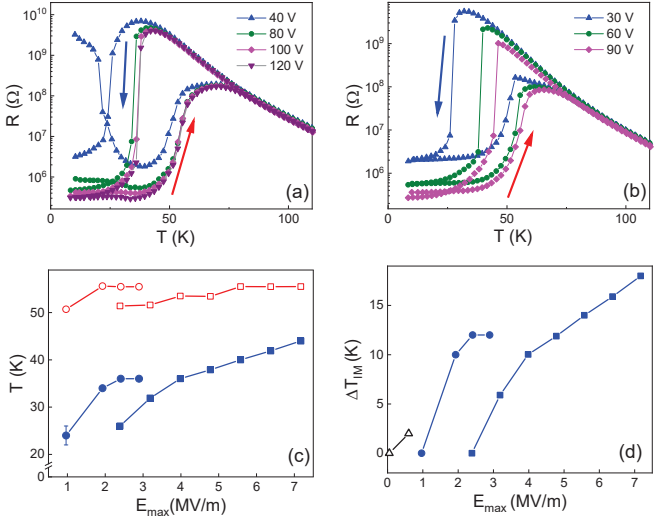}
    \caption{\label{rtemp} $R-T$ curves for an (La$_{0.4}$Pr$_{0.6}$)$_{0.67}$Ca$_{0.33}$MnO$_3$ thin film grown on (110) NdGaO$_3$ measured with a non-uniform electric field produced by (a) pattern 1 and (b) pattern 2. The arrows in both figures show the direction of temperature change. (c) $T-E_{\mathrm{max}}$ plots where, blue filled circles and squares represent $T_{\mathrm{IM}}$ for patterns 1 \& 2, respectively and red open circles and squares represent $T_{\mathrm{MI}}$ for patterns 1 \& 2, respectively.(d) Plots of $\Delta T_{\mathrm{IM}}$ vs. $E_{\mathrm{max}}$ for pattern 1 (blue filled circles) and pattern 2 (blue filled squares). The open triangles indicate $\Delta T_{\mathrm{IM}}$ for uniform applied electric field (data taken from Ref. \onlinecite{Jeen}).}
\end{figure}

\section{Results and Discussions}

The electrode patterns shown in Figs. 1(c) and (d) were chosen as two simple configurations for producing non-uniform electric fields:  a circular configuration (pattern 1) for which the electric field is given by Eq. 2 
\begin{equation}
\abs{\boldsymbol{E}} = \frac{V_A}{r \ln (r_1/r_2)}
\end{equation}
($V_{\mathrm{A}}$ is the applied voltage difference, $r_1$ and $r_2$ are radii of the inner and outer electrodes respectively, $r$ is the distance from the inner electrode center) and pattern 2 which is essentially a sharp projection separated from a straight conductor. Figs. 1(c) and (d) also show the electric field magnitude distribution for each pattern calculated using the relaxation technique. The in-plane magnetic anisotropy of the LPCMO thin films introduces an additional factor in the design of the electrodes as will be discussed next. Since the electric field direction for pattern 1 is radial, both the magnetic easy ([1$\overline{1}$0] direction) and hard axis ([001] direction) are subjected to the same electric field. Under such conditions, the FMM regions in LPCMO show a preference for aligning along the easy axis as was reported in Ref. \onlinecite{Jeen} where, it was also shown that even without an applied field the FMM regions tend to align along the easy axis. Hence, we chose the specific shape and orientation of pattern 2 to apply the highest electric field and maximum $\nabla\abs{\boldsymbol{E}}^2$ along the hard axis to isolate just the electric field effect on the dynamic percolation of the FMM regions. The shape of pattern 2 also avoids sharp corners which could provide alternate regions of dynamic percolation and a slight asymmetry ensures that the strongest electric field (and maximum $\nabla\abs{\boldsymbol{E}}^2$) is in the region shown by the dotted circle in Fig. ~\ref{sample}(d). The maximum electric field ($E_{\mathrm{max}}$) for patterns 1 and 2 are (2.4 $\times 10^4$)$V_{\mathrm{A}}$ V/m and (8.0 $\times 10^4$)$V_{\mathrm{A}}$ V/m respectively, where $V_{\mathrm{A}}$ is the voltage applied across the two electrodes.

Fig. ~\ref{rtemp}(a) shows the voltage dependent $R-T$ plots for LPCMO using pattern 1.  The resistance was measured using a two-probe method with a constant voltage applied across the electrodes shown in Fig. ~\ref{sample}(c). Hence, there is a question about the contribution of the Au/Ti contacts to the overall resistance. To check for this contribution, we calculated the resistivity of LPCMO using the equation for the geometry of pattern 1: 
\begin{equation}
\rho = R\frac{2\pi d}{\ln(r_1/r_2)}
\end{equation}
where $d$ is the film thickness. A comparison of the calculated resistivity values to those reported in Refs. \onlinecite{Jeen,Dhakal} shows that the contribution of contact resistance is negligible for these two probe measurements. The hysteresis observed between the cooling and warming runs of the $R-T$ plot is expected due to phase separation in LPCMO and is dependent on the thermal history of the sample. In Fig. ~\ref{rtemp}(a), each $R-T$ measurement for pattern 1 was started at 10 K and the sample was then heated up to 200 K in steps of 2 K with $V_{\mathrm{A}}$ turned on followed by a cooling run from 200 K to 8 K. No insulator-to-metal transition was observed for $V_{\mathrm{A}} < 40$ V. The anomalous high resistance of about 3 G$\Omega$, observed below 20 K in the warming run at 40 V, drops to about 1 M$\Omega$ on warming. This behavior has been reported before in LPCMO microstructures and is due to ``frozen" FMM regions at low temperatures which become unblocked on warming and form a conducting path in the presence of an electric field ~\cite{guneeta1,delozanne1,cheongsharma,ghivelder}. 

A preliminary inference from Fig. ~\ref{rtemp}(a) is that the non-uniform electric field produces a larger increase in the insulator-to-metal transition temperature ($\approx$ 12 K) compared to a uniform field ($\approx$ 2 K) ~\cite{Jeen}. To discuss this effect quantitatively, we define the insulator-to-metal/metal-to-insulator transition temperature ($T_{\mathrm{IM}}$/$T_{\mathrm{MI}}$) as the temperature at which the slope of the $\log(R)-T$ curve is maximum in the cooling/warming run. Since the electric field is non-uniform, we will use the maximum value of the electric field ($E_{\mathrm{max}}$) for a given $V_{\mathrm{A}}$ to compare the results to the case of a uniform electric field. Fig. ~\ref{rtemp}(c) shows the variation of $T_{\mathrm{IM}}$ and $T_{\mathrm{MI}}$ as a function of $E_{\mathrm{max}}$ (using $E_{\mathrm{max}}=(2.4 \times 10^4)V_{\mathrm{A}}$ V/m for pattern 1). $T_{\mathrm{IM}}$ increases with increasing $E_{\mathrm{max}}$ and plateaus at about 36 K. We then defined the change in $T_{\mathrm{IM}}$ relative to its value for $V_{\mathrm{A}}$ = 40 V as $\Delta T_{\mathrm{IM}} = T_{\mathrm{IM}}(V)-T_{\mathrm{IM}}(40 \mathrm{V})$ for pattern 1, which is plotted as function of $E_{\mathrm{max}}$ in Fig. ~\ref{rtemp}(d). $\Delta T_{\mathrm{IM}}$ plateaus at about 12 K for pattern 1. To illustrate the difference between the effects of a non-uniform and uniform applied electric fields, Fig. ~\ref{rtemp}(d) also shows the $\Delta T_{\mathrm{IM}}$ (defined as $T_{\mathrm{IM}}(V)-T_{\mathrm{IM}}(5 \mathrm{V})$) for the device structure discussed in Ref. \onlinecite{Jeen} (open triangles). The $\Delta T_{\mathrm{IM}}$ of 12 K due to the non-uniform electric field produced by pattern 1 is significantly larger than the $\Delta T_{\mathrm{IM}}$ of about 2 K due to the uniform electric field~\cite{Jeen}. However, $\Delta T_{\mathrm{IM}}$ does top out at 12 K for pattern 1 which may be due to the accumulation of the FMM regions near the inner electrode where the electric field is maximum. 

To mitigate this issue with pattern 1, we then switched to pattern 2 where the distance between the electrodes is smaller than pattern 1 (see Fig. ~\ref{sample}). In addition, the maximum $\nabla\abs{\boldsymbol{E}}^2 = 1.7 \times10^{17}$ V$^2$/m$^3$ for pattern 2 which is larger than that of the pattern 1 (maximum $\nabla\abs{\boldsymbol{E}}^2 = 1.6 \times10^{16}$ V$^2$/m$^3$) by an order of magnitude. The $R-T$ curves at different $V_{\mathrm{A}}$ measured using pattern 2 are shown in Fig. ~\ref{rtemp}(b). Figs. ~\ref{rtemp}(c) and ~\ref{rtemp}(d) show that the increase in $T_{\mathrm{IM}}$ as a function of $E_{\mathrm{max}}$ is larger for pattern 2 compared to pattern 1. In addition, the plateauing of $\Delta T_{\mathrm{IM}}$ as a function of $E_{\mathrm{max}}$ observed for pattern 1 is absent for pattern 2. It is also notable that while a uniform field left $T_{\mathrm{MI}}$ unchanged~\cite{Jeen}, for both patterns 1 and 2 a small increase of about 4 K in $T_{\mathrm{MI}}$ was observed as function of $E_{\mathrm{max}}$ in the warming cycle (represented by open circles for pattern 1 and open squares for pattern 2 in Fig. ~\ref{rtemp}(c)). The rise in $T_{\mathrm{MI}}$ is unexpected since the electric field should not affect the low temperature static phase separated (SPS) state during warming run due to pinned phase boundaries ~\cite{Dhakal}. A likely scenario is that the stronger alignment of the FMM regions by the non-uniform electric field in the FPS state leads to a ``frozen-in" alignment of the FMM regions when the sample is cooled to the SPS state. Subsequently, when the sample is warmed up, the ``frozen-in" alignment leads to a higher $T_{\mathrm{MI}}$.
\begin{figure}
\includegraphics[width=\columnwidth]{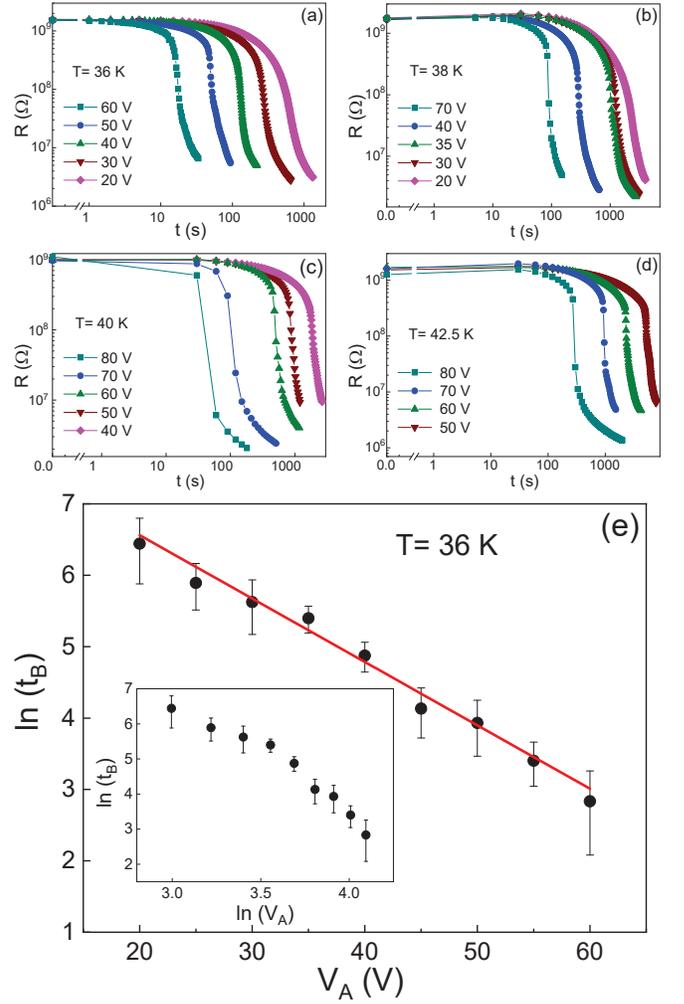}
\caption{\label{rtime1} (a), (b), (c), \& (d) Voltage dependence of $R-t$ curves for pattern 1 at 36 K, 38 K, 40 K, \& 42.5 K, respectively. (e) Semilog plot of $t_{\mathrm{B}}$ as a function of $V_{\mathrm{A}}$ at 36 K. Solid red line is a linear fit to the data with fitting parameters (see Eq. 4) $A= (4\times10^3\pm10^3) s$ and $k$ = ($0.09\pm 0.01$) $V^{-1}$. Inset: The corresponding ln-ln plot of $t_{\mathrm{B}}$ vs. $V_{\mathrm{A}}$ at 36 K.}
\end{figure}

\begin{figure}
\includegraphics[width=\columnwidth]{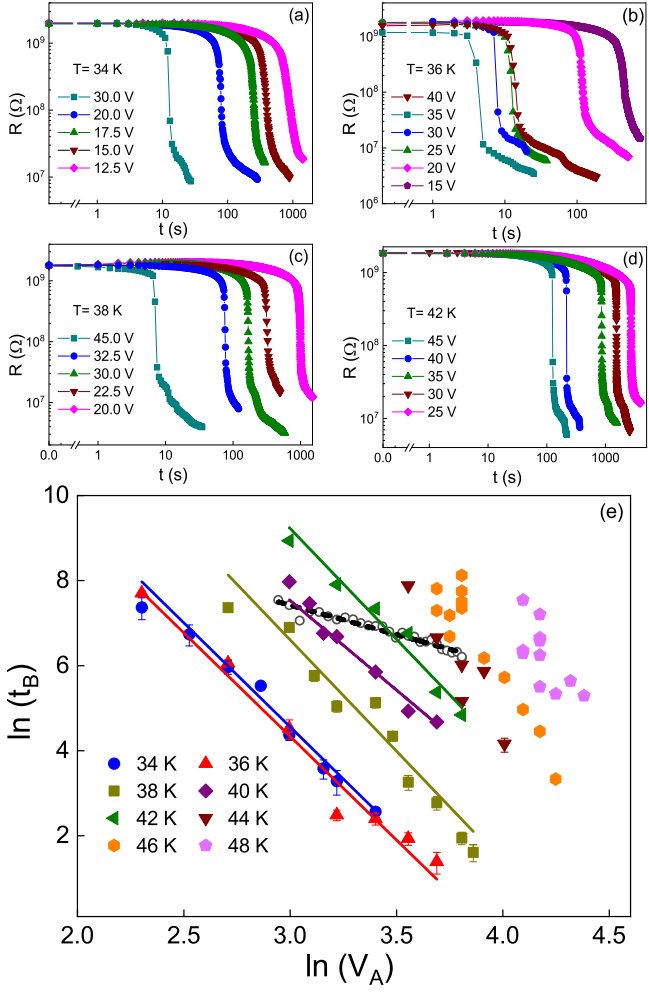}
    \caption{\label{rtime2} (a), (b), (c), \& (d) Voltage dependent $R-t$ curves for pattern 2 at 34 K, 36 K, 38 K, \& 42 K, respectively. (e) Ln-ln plots depicting $t_{\mathrm{B}}$ as a function of $V_{\mathrm{A}}$ for the temperature range of 34 K to 48 K. The solid lines are the corresponding linear fits for the data from 34 K to 42 K. The open black circles represent the data points for a uniform electric field (data taken from Ref. \onlinecite{Jeen}) and the corresponding linear fit (black dotted line).}
\end{figure}

While the $R-T$ measurements discussed above provide the initial evidence of a stronger dielectrophoretic effect due to non-uniform electric fields, in Ref. \onlinecite{Jeen} it was resistance vs. time ($R-t$) measurements that had provided the clearest signature of dynamic percolation. A sharp drop was observed in the resistance after a certain amount of time in an applied electric field ~\cite{Jeen}. It was also shown in Ref. \onlinecite{Jeen} that the time required for this resistance breakdown (breakdown time, $t_{\mathrm{B}}$) decreased exponentially with an increase in the applied (uniform) electric field along the magnetic easy axis whereas, when the electric field was applied along the hard axis, the decrease in $t_{\mathrm{B}}$ with electric field was non-exponential (and will be shown to follow a power law in this paper). A non-uniform electric field is expected to have a stronger effect on the FMM regions since the dielectrophoretic force is proportional to $\nabla\abs{\boldsymbol{E}}^2$ and also due to the introduction of higher order multipole moments and field derivatives~\cite{Nili}. Hence, we probed the dynamic behavior of the FPS state in a non-uniform electric field to further distinguish the effects from a uniform electric field. We performed isothermal resistance measurements as a function of time ($R-t$) using patterns 1 and 2 at different temperatures and $V_{\mathrm{A}}$. Figs. 3(a) to (d) show the $R-t$ curves for pattern 1. These temperatures are above the lower limit of $T_{\mathrm{IM}}$ for pattern 1 and therefore, below the percolation threshold of the fluidlike FMM phase ~\cite{Zhang} such that the sample remains in the insulating state unless a certain threshold voltage is applied for a certain period of time $t_{\mathrm{B}}$ ~\cite{Jeen, Dhakal}. Once the resistance breakdown occurs, it remains at the low value even when the electric field is removed. Therefore, the sample needs to be heated to a temperature well above the $T_{\mathrm{MI}}$ and then cooled to the next desired temperature for another isothermal $R-t$ measurement~\cite{Jeen,Dhakal}. We heated the sample to 150 K before each $R-t$ measurement. We also kept the rate of cooling consistent so that the initial resistance of the sample at time $t=0$ is same for each voltage dependent $R-t$ measurement.

To study the behavior of $t_{\mathrm{B}}$ as a function of $V_{\mathrm{A}}$ for pattern 1, we first defined $t_{\mathrm{B}}$ as the time at which the slope of the $\log(R)-t$ curve (not shown), at a particular temperature and $V_{\mathrm{A}}$, is maximum. Error bars were calculated using the full width at half maximum (FWHM) of the derivative of $\log(R)$ with respect to $t$. Fig. ~\ref{rtime1}(e) shows the $\ln(t_{\mathrm{B}})$ vs. $V_{\mathrm{A}}$ plot at 36 K.  The red curve is a straight line fit showing the exponential behavior of $t_{\mathrm{B}}$ as a function of $V_{\mathrm{A}}$. Such an exponential behavior was also reported in Ref. \onlinecite{Jeen} when the electric field was applied along the easy axis and it was shown that  $t_{\mathrm{B}}$ and $V_{\mathrm{A}}$ follow the relation:
\begin{equation}
t_{\mathrm{B}} = A \exp(-kV_{\mathrm{A}})
\end{equation}
where $k$ and $A$ are fitting parameters (fitted values at 36 K given in Fig. ~\ref{rtime1} caption). The radial electric field produced by pattern 1 implies that an equal electric field (and $\nabla\abs{\boldsymbol{E}}^2$) is applied along both the easy and hard axes. Since the FMM regions preferentially align along the easy axis, the $R-t$ behavior is dominated by the dynamic percolation of the FMM regions along the easy axis for pattern 1 which leads to the behavior of $t_{\mathrm{B}}$ described by Eq. 4. The inset of Fig. ~\ref{rtime1}(e) shows the same data as a ln-ln plot to illustrate the lack of a power law relation between $t_{\mathrm{B}}$ and $V_{\mathrm{A}}$. The decrease of $t_{\mathrm{B}}$ with $V_{\mathrm{A}}$ according to Eq. 4 was also observed at other temperatures except at 38 K. This exponential behavior was unexpected since earlier studies on flocculation of particles in an external electric field predicted a power law behavior i.e.:
\begin{equation}
t_{\mathrm{B}} \propto V_{\mathrm{A}}^{-\delta}
\end{equation}
where $\delta = 2$ for the particular case of dipolar interactions in a uniform electric field~\cite{Tang}. It is likely that the tendency of the FMM regions to align along the easy axis even in the absence of an electric field leads to the deviation from the power law behavior of Eq. 5 to the exponential behavior of Eq. 4. Indeed, in Ref. \onlinecite{Jeen} it was shown that when the electric field was applied along the hard axis, $t_{\mathrm{B}}$ and $V_{\mathrm{A}}$ did not follow an exponential relation. Hence, to extract the effect purely due to a non-uniform electric field (without the influence of the magnetic interactions as in along the easy axis) we used pattern 2 which is expected to constrain the motion of the FMM regions along the hard axis. 

We measured the dynamic behavior of the FMM regions using pattern 2 and Figs. 4(a) to (d) show the voltage dependent $R-t$ curves at different temperatures. Fig. ~\ref{rtime2}(e) shows the $\ln(t_{\mathrm{B}})$ vs. $\ln(V_{\mathrm{A}})$ plots for the different temperatures. The straight line fits to these ln-ln plots show that for the dynamic percolation along the hard axis there is indeed a power law relation between $t_{\mathrm{B}}$ and $V_{\mathrm{A}}$ (Eq. 5) and not an exponential behavior. The exponents $\delta$ measured at different temperatures are $4.9 \pm 0.3$ (at 34 K), $4.9\pm 0.3$ (at 36 K), $5.2\pm 0.5$ (at 38 K), $4.2\pm 0.3$ (at 40 K), and $5.2\pm 0.6$ (at 42 K) which are significantly different from $\delta = 2$ (expected value for uniform electric field ~\cite{Tang}). We claim that this deviation from the dipolar approximation is due to the non-uniform external field. To bolster this claim, the $\ln(t_{\mathrm{B}})$ versus $\ln(V_{\mathrm{A}})$ for a uniform external field (taken from Ref. \onlinecite{Jeen}) is also plotted in Fig. ~\ref{rtime2}(e). $\delta =1.37 \pm 0.06$ for the case of a uniform external field. The reduction of $\delta$ to a value less than 2 has been observed before and is possibly due to screening of the electric field due to the metallicity of the FMM regions ~\cite{Tang}. Nevertheless, it is clear that a non-uniform electric field leads to a significantly stronger power law behavior of $t_{\mathrm{B}}$ as a function of $V_{\mathrm{A}}$ compared to a uniform external field. Fig. ~\ref{rtime2}(e) also shows that for the higher temperatures of 44 K, 46 K \& 48 K, the scatter in the data does not allow a straight line fit suggesting a breakdown of the power law behavior possibly due to the smaller FMM regions with a wider variation in size, at higher temperatures.

\begin{figure}
\includegraphics[width=\columnwidth]{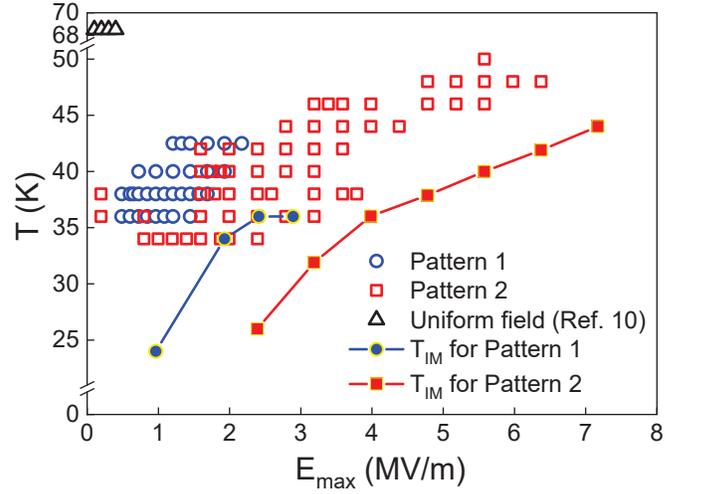}
   \caption{\label{phasediag} $T$ vs. $E_{\mathrm{max}}$ phase diagram showing the region in which dynamic percolation is observed for pattern 1 (open blue circles), pattern 2 (open red squares), and uniform electric field (open black triangles, data taken from Ref. \onlinecite{Jeen}). The $T_{\mathrm{IM}}$ for different $E_{\mathrm{max}}$ are also reproduced here for pattern 1 (solid blue circles) and pattern 2 (solid red squares).}
\end{figure}

\section{Conclusions}

A non-uniform electric field has a stronger effect on the dynamic percolation of FMM regions in phase separated LPCMO compared to a uniform field which reinforces the argument in favor of the dielectrophoresis model for dynamic percolation. Fig. ~\ref{phasediag} summarizes the results as a temperature-electric field phase diagram showing the region where dynamic percolation is observed which, for both pattern 1 (open blue circles) and 2 (open red squares), is much larger when the electric field is non-uniform compared to the case of a uniform field (open black triangles). The breakdown (percolation) time $t_{\mathrm{B}}$ in the magnetic hard axis direction and the applied voltage $V_{\mathrm{A}}$ follow a power law relation as in Eq. 5 with $\delta \approx 5$ which is higher than the case of a uniform field. Such a high value of $\delta$ in a non-uniform field is expected for non-spherical FMM regions of length scale similar to that of the electric field non-uniformity~\cite{Nili}. Our results suggest a unique method for controlling micrometer sized ferromagnetic regions using electric fields and simple electrode configurations. Direct imaging of the phase separation in LPCMO in the presence of non-uniform electric fields using techniques such as magnetic force microscopy are needed to confirm the dielectrophoresis model for dynamic percolation.

\section{Data availability statement}

The data that supports the findings of this study are available within the article

\begin{acknowledgments}
This work was supported by NSF DMR-1410237. We thank M. Meisel for magnetization measurements.

\end{acknowledgments}

\end{document}